\documentclass[twocolumn,superscriptaddress,showpacs,floatfix,prc]{revtex4}
\usepackage{graphicx}

\begin{document}

\title{Isospin effects on two-nucleon correlation functions in\ heavy-ion
collisions at intermediate energies}
\author{Lie-Wen Chen}
\thanks{On leave from Department of Physics, Shanghai Jiao Tong University,
Shanghai 200030, China}
\affiliation{Cyclotron Institute and Physics Department, Texas A\&M University, College
Station, Texas 77843-3366}
\author{V. Greco}
\affiliation{Cyclotron Institute and Physics Department, Texas A\&M University, College
Station, Texas 77843-3366}
\author{C. M. Ko}
\affiliation{Cyclotron Institute and Physics Department, Texas A\&M University, College
Station, Texas 77843-3366}
\author{Bao-An Li}
\affiliation{Department of Chemistry and Physics, P.O. Box 419, Arkansas State
University, State University, Arkansas 72467-0419}
\date{\today}

\begin{abstract}
Using an isospin-dependent transport model, we study isospin effects on
two-nucleon correlation functions in heavy-ion collisions induced by
neutron-rich nuclei at intermediate energies. We find that these correlation
functions are sensitive to the density dependence of nuclear symmetry
energy, but not to the incompressibility of symmetric nuclear matter and the
medium dependence of nucleon-nucleon cross sections. This sensitivity is
mainly due to effects of nuclear symmetry energy on the emission times of
neutrons and protons as well as their relative emission sequence. We also
study the variations of the symmetry energy effects on nucleon-nucleon
correlations with respect to the impact parameter, incident energy, and mass
number of heavy ion collisions.
\end{abstract}

\maketitle

\section{Introduction}

Heavy ion collisions with radioactive nuclear beams provide a unique
opportunity to study the properties of nuclear matter at the extreme
condition of large isospin asymmetry \cite{npa01,ibook,li98,ditoro99}. In
these collisions, the dynamics is affected not only by the
isospin-independent part of nuclear equation of state (\textrm{EOS}) but
also by the isospin-dependent part, i.e., the nuclear symmetry energy $E_{%
\mathrm{sym}}(\rho )$, and the isospin-dependent in-medium nucleon-nucleon (%
\textsl{N}-\textsl{N}) cross sections. Knowledge on the density dependence
of nuclear symmetry energy is essential for understanding both the structure
of radioactive nuclei \cite{oya,brown,hor01,furn02} and many key issues in
astrophysics \cite{bethe,lat01,bom}. Although the root-mean-squared radius
of neutron distribution in a heavy nucleus and that of a neutron star differ
by about $18$ orders of magnitude, the properties of both systems are
strongly affected by the same $E_{\mathrm{sym}}(\rho )$ \cite{hor01,eng,pra}%
. Moreover, the nucleosynthesis in pre-supernova evolution of massive stars,
mechanisms of supernova explosions, cooling rates of protoneutron stars and
associated neutrino fluxes, and the kaon condensation as well as the hadron
to quark-gluon plasma phase transitions in neutron stars all rely critically
on the nuclear symmetry energy \cite{bethe,lat01,ditoro}. However, our
knowledge on $E_{\mathrm{sym}}(\rho )$ is still rather poor despite
extensive theoretical studies based on various many-body theories \cite{bom}%
. In fact, it is often regarded as the most uncertain property of a
neutron-rich matter \cite{kut,kub02}. Fortunately, radioactive beams,
particularly the very energetic ones to be available at the planned rare
isotope accelerator (RIA) and the new accelerator facility at German Heavy
Ion Accelerator Center (GSI), provide a great opportunity to study the $E_{%
\mathrm{sym}}(\rho )$. It is thus important to investigate theoretically
what experimental observables can be used to extract information about $E_{%
\mathrm{sym}}(\rho )$. In this regard, significant progress has been made
recently. For instance, measurement of neutron skins of radioactive nuclei
via their total reaction cross sections \cite{oya} or that in stable heavy
nuclei, such as $^{208}$Pb, via parity-violating electron scatterings \cite%
{hor01} was recently proposed as a sensitive probe to $E_{\mathrm{sym}}(\rho
)$. In heavy-ion collisions induced by neutron-rich nuclei, the
pre-equilibrium neutron/proton ratio \cite{li97}, isospin fractionation \cite%
{fra1,fra2,xu00,tan01,bar02}, isoscaling in multifragmentation \cite{betty},
proton differential elliptic flow \cite{lis}, neutron-proton differential
transverse flow \cite{li00} as well as the $\pi ^{-}$ to $\pi ^{+}$ ratio %
\cite{li02} have been identified as promising probes.

Observables mentioned in the above involve only the momentum, charge, and
mass distributions of individual particles in the final state of a reaction.
On the other hand, two-particle correlation functions, through final-state
interactions and quantum statistics effects, have been shown to be a
sensitive probe to the temporal and spatial information about the reaction
dynamics in heavy ion collisions at incident energies ranging from
intermediate \cite{Boal90,bauer,ardo97} to RHIC energy \cite{Wied99}. In
particular, the correlation function of two nonidentical particles from
heavy ion collisions has been found to depend on their relative space-time
distributions at freeze out and thus provides a useful tool for measuring
the emission sequence, time delay, and separation between the emission
sources for different particles \cite%
{Geld95,Lednicky96,Voloshin97,Ardouin99,Pratt99,Gourio00}. Indeed, time
delays as short as $1$ \textrm{fm/c} and source radius differences of a few
tenth \textrm{fm} have been resolved experimentally from correlation
functions of nonidentical particles \cite{Kotte99}.

In the present work, we study systematically how $E_{\mathrm{sym}}(\rho )$
affects the temporal and spatial structure of reaction dynamics in heavy-ion
collisions induced by neutron-rich nuclei at intermediate energies. A brief
report of present work can be found in Ref. \cite{npCor}. We show that
average emission times of neutrons and protons as well as their relative
emission sequence in heavy ion collisions are sensitive to $E_{\mathrm{sym}%
}(\rho )$. A stiffer density dependence of nuclear symmetry energy leads to
faster and nearly simultaneous emissions of high momentum neutrons and
protons. Consequently, strengths of the correlation functions for nucleon
pairs with high total momentum, especially neutron-proton pairs with low
relative momentum, are stronger for a stiffer $E_{\mathrm{sym}}(\rho )$.
This novel property thus provides another possible tool for extracting
useful information about $E_{\mathrm{sym}}(\rho )$. In addition, we also
study how two-nucleon correlation functions depend on the
isospin-independent part of nuclear matter \textrm{EOS}, the in-medium 
\textsl{N}-\textsl{N} cross sections as well as the impact parameter,
incident energy, and masses of colliding nuclei.

The paper is organized as follows. In Section \ref{symmetry}, we discuss the
density dependence of nuclear symmetry energy and the \textrm{EOS} of
neutron-rich nuclear matter. The Isospin-dependent
Boltzmann-Uehling-Uhlenbeck (IBUU) transport model used in present study is
briefly described in Section \ref{ibuu}. In Section \ref{source}, we
presented results from the IBUU model on the nucleon emission functions and
their isospin dependence in heavy-ion collisions induced by neutron-rich
nuclei at intermediate energies. These include the nucleon emission times,
the size of nucleon emission source, and the momentum distributions of
emitted nucleons. In Section \ref{correlation}, we study nucleon-nucleon
correlation functions and their dependence on the nuclear symmetry energy.
We also present results on the variations of the symmetry energy effects on
nucleon-nucleon correlations with respect to the nuclear isoscalar
potential, nucleon-nucleon cross sections, as well as the impact parameter,
incident energy, and masses of heavy ion collisions. Finally, we conclude
with a summary and outlook in Section \ref{summary}.

\section{Nuclear symmetry energy}

\label{symmetry}

Many theoretical studies (e.g., \cite%
{wir88,lat91,siemens70,baym71,prak88,thor94}) have shown that the \textrm{EOS%
} of asymmetric nuclear matter can be approximately expressed as 
\begin{equation}
E(\rho ,\delta )=E(\rho ,\delta =0)+E_{\text{\textrm{sym}}}(\rho )\delta
^{2},  \label{eq1}
\end{equation}
where $\rho =\rho _{n}+\rho _{p}$ is the baryon density; $\delta
=(\rho_{n}-\rho _{p})/(\rho _{p}+\rho _{n})$ is the isospin asymmetry; and $%
E(\rho,\delta =0)$ is the energy per particle in symmetric nuclear matter.
The bulk symmetry energy is denoted by the so-called symmetry energy
coefficient $E_{\text{sym}}(\rho )\equiv E(\rho ,\delta =1)-E(\rho ,\delta
=0)$. Its value at normal nuclear matter density $\rho _{0}$, i.e., $E_{%
\text{sym}}(\rho _{0})$, is predicted to be about $34\pm 4$ \textrm{MeV} and
is comparable to the value extracted from atomic mass data \cite{mass}.

The symmetry energy coefficient $E_{\text{\textrm{sym}}}(\rho )$\ can be
expanded around normal nuclear matter density as 
\begin{equation}
E_{\text{\textrm{sym}}}(\rho )=E_{\text{\textrm{sym}}}(\rho _{0})+\frac{L}{3}%
\left( \frac{\rho -\rho _{0}}{\rho _{0}}\right) +\frac{K_{\text{\textrm{sym}}%
}}{18}\left( \frac{\rho -\rho _{0}}{\rho _{0}}\right) ^{2},  \label{eq2}
\end{equation}%
where $L$ and $K_{\text{\textrm{sym}}}$ are the slope and curvature of
symmetry energy coefficient at normal nuclear density, i.e., 
\begin{eqnarray}
L &=&3\rho _{0}\frac{\partial E_{\text{\textrm{sym}}}(\rho )}{\partial \rho }%
|_{\rho =\rho _{0}},  \label{L} \\
K_{\text{\textrm{sym}}} &=&9\rho _{0}^{2}\frac{\partial ^{2}E_{\text{\textrm{%
sym}}}(\rho )}{\partial ^{2}\rho }|_{\rho =\rho _{0}}.  \label{eq4}
\end{eqnarray}%
The $L$ and $K_{\text{sym}}$ characterize the density dependence of nuclear
symmetry energy around normal nuclear matter density, and thus provide
important information about the properties of nuclear symmetry energy at
both high and low densities. They are, however, not well determined either
theoretically or experimentally. The theoretically predicted values for $K_{%
\mathrm{sym}}$ vary from about $-700$ \textrm{MeV} to $+466$ \textrm{MeV}
(e.g. \cite{bom91}), while data from giant monopole resonances indicate that 
$K_{\text{sym}}$ is in the range from $-566\pm 1350$ MeV to $34\pm 159$ 
\textrm{MeV} depending on the mass region of nuclei and the number of
parameters used in parameterizing the incompressibility of finite nuclei %
\cite{shl93}. There are no experimental data for the value of $L$, and
different theoretical models give its value from $-50$ to $200$ \textrm{MeV} %
\cite{furn02}.

Predictions for the density dependence of $E_{\mathrm{sym}}(\rho )$ are thus
divergent. In the present study, we adopt the parameterization used in Ref. %
\cite{hei00} for studying the properties of neutron stars, i.e., 
\begin{equation}
E_{\mathrm{sym}}(\rho )=E_{\mathrm{sym}}(\rho _{0})\cdot u^{\gamma },
\label{srho2}
\end{equation}%
where $u\equiv \rho /\rho _{0}$ is the reduced density and $E_{\mathrm{sym}%
}(\rho _{0})=35$ MeV is the symmetry energy at normal nuclear matter
density. The symmetry potential acting on a nucleon derived from the above
nuclear symmetry energy is \cite{lis} 
\begin{eqnarray}
V_{\mathrm{sym}}(\rho ,\delta ) &=&\pm 2[E_{\mathrm{sym}}(\rho
_{0})u^{\gamma }-12.7u^{2/3}]\delta  \nonumber  \label{vasy} \\
&&+[E_{\mathrm{sym}}(\rho _{0})(\gamma -1)u^{\gamma }+4.2u^{2/3}]\delta ^{2},
\end{eqnarray}
where ``+'' and ``--'' are for neutrons and protons, respectively.

For the isospin-independent part of the nuclear \textrm{EOS}, $E(\rho
,\delta =0)$, we use a Skyrme-like parameterization, i.e., 
\begin{equation}
E(\rho ,\delta =0)=\frac{3}{5}E_{F}^{0}u^{2/3}+\frac{a}{2}u+\frac{b}{%
1+\sigma }u^{\sigma },  \label{EOSsky}
\end{equation}%
where the first term is the kinetic part and $E_{F}^{0}$ is the Fermi energy
of symmetric nuclear matter at normal nuclear matter density. The parameters 
$a=-358.1$ \textrm{MeV}, $b=304.8$ \textrm{MeV}, and $\sigma =7/6$
correspond to the so-called soft \textrm{EOS} with incompressibility $%
K_{0}=201$ \textrm{MeV}, while $a=-123.6$ \textrm{MeV}, $b=70.4$ \textrm{MeV}%
, and $\sigma =2$ give the so-called stiff \textrm{EOS} with
incompressibility $K_{0}=380$ \textrm{MeV}. Both EOS's have a saturation
density $\rho _{0}=0.16$ fm$^{-3}$ for symmetric nuclear matter. The
isospin-independent potential obtained from this \textrm{EOS} is then 
\begin{equation}
V_{0}(\rho )=au+bu^{\sigma }.
\end{equation}

The pressure of nuclear matter is given by 
\begin{equation}
P(\rho ,\delta )=\rho ^{2}\frac{\partial E(\rho ,\delta )}{\partial \rho },
\end{equation}%
which leads to a contribution from nuclear symmetry energy given by 
\begin{eqnarray}
P_{\text{\textrm{sym}}}(\rho ,\delta ) &=&\rho ^{2}\delta ^{2}\frac{\partial
E_{\mathrm{sym}}(\rho )}{\partial \rho }  \nonumber \\
&=&\rho _{0}E_{\mathrm{sym}}(\rho _{0})\gamma \delta ^{2}u^{\gamma +1},
\end{eqnarray}%
besides that due to isospin-independent energy given by 
\begin{equation}
P_{\mathrm{0}}(\rho ,\delta =0)=\frac{2}{5}E_{F}^{0}\rho _{0}u^{5/3} +\frac{1%
}{2}a\rho _{0}u^{2}+\frac{b\sigma }{1+\sigma }u^{\sigma +1}.
\end{equation}

At moderate and high temperatures $T$ ($T>4$ \textrm{MeV}), the pressure can
also be decomposed into an isospin-dependent part $P_{\text{\textrm{sym}}%
}(\rho ,\delta ,T)$ and an isospin-independent part $P_{\text{\textrm{0}}%
}(\rho ,\delta =0,T)$, i.e., 
\begin{eqnarray}
P_{\text{\textrm{sym}}}(\rho ,\delta ,T) &=&\frac{1}{2}\sum_{n=1}^{\infty
}b_{n}\left( \frac{\lambda _{T}^{3}\rho }{4}\right) ^{n}[(1+\delta )^{1+n} 
\nonumber  \label{eqPsymT} \\
&+&(1-\delta )^{1+n}-2]+\rho _{0}\delta ^{2}  \nonumber \\
&\times &\left[ E_{\mathrm{sym}}(\rho _{0})\gamma u^{\gamma +1}-8.5u^{5/3}%
\right] ,
\end{eqnarray}%
and\bigskip 
\begin{eqnarray}
P_{\text{\textrm{0}}}(\rho ,\delta  &=&0,T)=T\rho \lbrack
1+\sum_{n=1}^{\infty }b_{n}(\frac{\lambda _{T}^{3}\rho }{4})^{n}]  \nonumber
\\
&&+\frac{1}{2}a\rho _{0}u^{2}+\frac{b\sigma }{1+\sigma }u^{\sigma +1},
\end{eqnarray}%
where $b_{n}^{\prime }$s are the inversion coefficients given in Ref. \cite%
{fra1,li98} and 
\begin{equation}
\lambda _{T}=\left( \frac{2\pi \hbar ^{2}}{mT}\right) ^{1/2}
\end{equation}%
is the thermal wavelength of nucleons with mass $m$. In Eq. (12), we have
used the explicit $\delta $-dependent kinetic part of symmetry energy
instead of the parabolic approximation to calculate the pressure at finite
temperatures. 
\begin{figure}[th]
\includegraphics[scale=0.8]{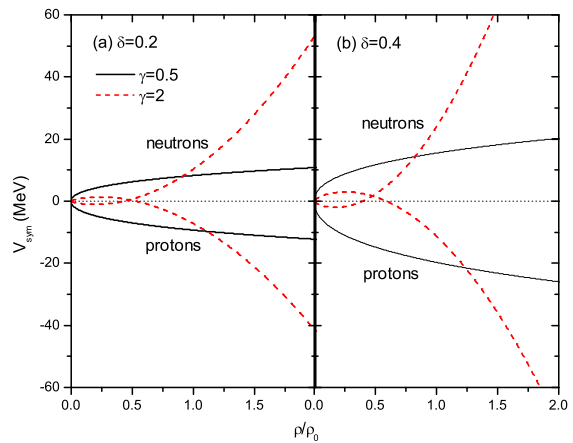} \vspace{0cm}
\caption{{\protect\small (Color online) Neutron and proton symmetry
potentials as functions of density in asymmetric nuclear matter with isospin
asymmetry (a) $\protect\delta =0.2$ and (b) $\protect\delta =0.4$ for both
stiff ($\protect\gamma =2)$ and soft ($\protect\gamma =0.5)$ symmetry
energies, respectively.}}
\label{SymPot}
\end{figure}

In the present work, we consider two cases of $\gamma =0.5$ (soft) and $2$
(stiff) to explore the large range of $L$ and $K_{\text{sym}}$ discussed
above. The $\gamma =0.5$ ($2$) gives $L=$ $52.5$ ($210.0$) \textrm{MeV }and $%
K_{\text{sym}}=-78.8$ ($630.0$) \textrm{MeV}. In Figs. \ref{SymPot} (a) and
(b), we show the nuclear symmetry potentials for protons and neutrons in
isospin asymmetric nuclear matter with $\delta =0.2$ and $0.4$,
respectively. Except at densities below $0.5\rho _{0}$ for the case of $%
\gamma =2$, the symmetry potential is attractive for protons and repulsive
for neutrons and increases with density and isospin asymmetry. Furthermore,
the stiff symmetry potential becomes stronger than the soft one when nuclear
density exceeds certain value around normal nuclear density, that depends on
the isospin asymmetry and is different for neutrons and protons. 
\begin{figure}[th]
\includegraphics[scale=1.1]{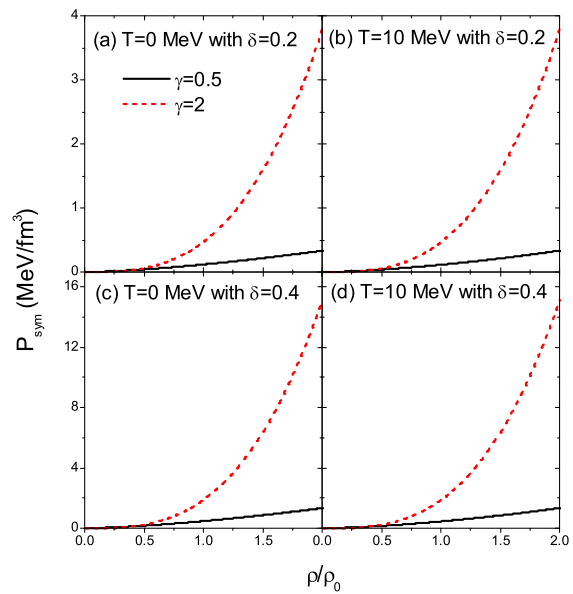} \vspace{0cm}
\caption{{\protect\small (Color online) Pressure of asymmetric nuclear
matter due to the symmetry energy, }$P_{\text{sym}}${\protect\small , as a
function of density for different temperatures and isospin asymmetries: (a) }%
$T=0${\protect\small \ MeV and }$\protect\delta =0.2,${\protect\small \ (b) }%
$T=10${\protect\small \ MeV and }$\protect\delta =0.2${\protect\small \
\thinspace\ (c) }$T=0${\protect\small \ MeV and }$\protect\delta =0.4$%
{\protect\small , and (d) }$T=10${\protect\small \ MeV and }$\protect\delta %
=0.4${\protect\small . Solid and dashed curves are for soft (}$\protect%
\gamma =2.0${\protect\small ) and stiff (}$\protect\gamma =0.5$%
{\protect\small ) symmetry energies, respectively.}}
\label{Psym}
\end{figure}
\begin{figure}[th]
\includegraphics[scale=0.8]{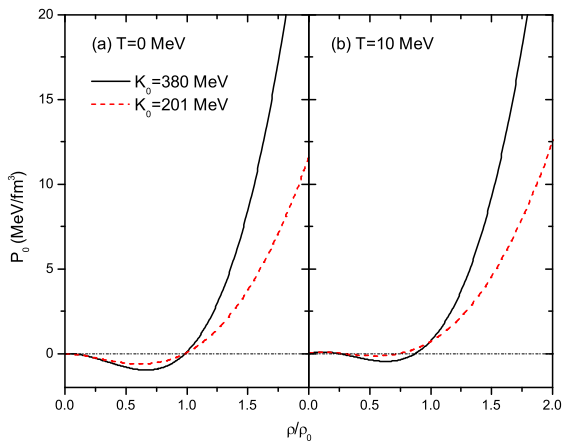} \vspace{0cm}
\caption{{\protect\small (Color online) Isospin-independent part of
pressure, }$P_{0}${\protect\small , as a function of density for }$T=0$%
{\protect\small \ (a) and }$10${\protect\small \ (b) MeV, respectively, with 
}$K_{0}{\protect\small =380}${\protect\small \ (solid curves) and }$%
{\protect\small 201}${\protect\small \ (dashed curves) MeV.}}
\label{P0}
\end{figure}

Different symmetry energies also result in different thermodynamical
properties for neutron-rich matter. The pressure due to symmetry energy, $P_{%
\mathrm{sym}}$, in an asymmetric nuclear matter with isospin asymmetry $%
\delta =0.2$ or $0.4$ and temperature $T=0$ or $10$ \textrm{MeV}, is shown
in Fig. \ref{Psym} as a function of density for both soft ($\gamma =0.5$)
and stiff ($\gamma =2$) symmetry energies. It is seen that the stiff
symmetry energy gives a larger pressure in asymmetric nuclear matter than
the soft one at same density, and the difference increases with both density
and isospin asymmetry. Since the symmetry energy is insensitive to
temperature, temperature dependence of the pressure from symmetry energy is
rather weak as in more sophisticated finite-temperature Skyrme-Hartree-Fock
calculations \cite{clw01}. For comparison, the isospin-independent part of
pressure, $P_{\text{\textrm{0}}}$, is shown in Fig. \ref{P0} as a function
of density for $T=0$ (a) and $10$ (b) \textrm{MeV}, and with $K_{0}=201$
(dashed curves) and $380$ (solid curves) \textrm{MeV}. We see that the stiff 
\textrm{EOS} gives a stronger pressure at both higher densities and
temperatures.

Although the total incompressibility of asymmetric nuclear matter is mainly
determined by the stiffness of the isoscalar part of nuclear energy, the
stiffness of symmetry energy has non-negligible effects on its \textrm{EOS}
and thermodynamical properties. We thus expect that different symmetry
energies would lead to different space-time evolutions in heavy-ion
collisions induced by neutron-rich nuclei at intermediate energies,
resulting in observable effects in the nucleon-nucleon correlation functions.

\section{Isospin-dependent Boltzmann-Uehling-Uhlenbeck (IBUU) transport model%
}

\label{ibuu}

\begin{figure}[th]
\includegraphics[scale=0.85]{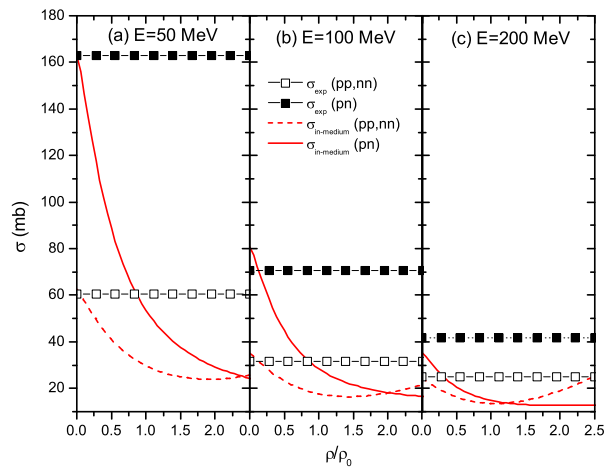} \vspace{0cm}
\caption{{\protect\small (Color online) Density dependence of total
nucleon-nucleon cross sections at laboratory energies }$E=50${\protect\small %
\ (a), } $100${\protect\small \ (b) and }$200${\protect\small \ (c) MeV, for
the experimental free-space cross sections (squares) and the in-medium
nucleon-nucleon cross sections from the Dirac-Brueckner approach based on
the Bonn A potential (lines).}}
\label{crsc}
\end{figure}

Our study is based on an isospin-dependent Boltzmann-Uehling-Uhlenbeck
(IBUU) transport model (e.g., \cite{buu,li97,li02,li00,li96}). In this
model, initial positions of protons and neutrons are determined according to
their density distributions from the relativistic mean-field (\textrm{RMF})
theory \cite{serot,zhu94,ren}. The neutron and proton initial momenta are
then taken to have uniform distributions inside their respective Fermi
spheres with Fermi momenta determined by their local densities via the
Thomas-Fermi approximation. For the isoscalar potential, we use as default
the Skyrme potential with incompressibility $K_{0}=380$ \textrm{MeV}.
Although the transverse flow data from heavy ion collisions are best
described by a momentum-dependent soft potential with $K_{0}=210$ \textrm{MeV%
} \cite{pan93,zhang93}, they can also be approximately reproduced with a
momentum-independent stiff potential with $K_{0}=380$ MeV. Dependence of the
reaction dynamics on isospins is included through the isospin-dependent
total and differential nucleon-nucleon cross sections and Pauli blockings,
the symmetry potential $V_{\mathrm{sym}}$, and the Coulomb potential $V_{%
\mathrm{c}}$ for protons. For nucleon-nucleon cross sections, we use as
default the experimental values in free space $\sigma _{\exp }$. For a
review of the IBUU model, we refer the reader to Ref. \cite{li98}. To study
effects due to the medium dependence of nucleon-nucleon cross sections, we
also use the parameterized in-medium nucleon-nucleon cross sections $\sigma
_{\text{in-medium}}$ from the Dirac-Brueckner approach based on the Bonn A
potential \cite{lgq9394}. In Figs. \ref{crsc} (a), (b) and (c), we show the
density dependence of total nucleon-nucleon cross sections $\sigma _{\exp }$
(squares) and $\sigma _{\text{\textrm{in-medium}}}$ (lines) for laboratory
incident energies $E=50$, $100$ and $200$ \textrm{MeV}, respectively. In
experimental free-space cross sections, the neutron-proton cross section is
about a factor of 3 larger than the neutron-neutron or proton-proton cross
section. On the other hand, in-medium nucleon-nucleon cross sections from
the Dirac-Brueckner approach based on the Bonn A potential have smaller
magnitude and weaker isospin dependence than $\sigma _{\exp }$ but strong
density dependence. 

To solve the IBUU model, we use the usual test particle method \cite{buu}.
Results presented in the following are obtained with $10,000$ events using $%
100$ test particles for a physical nucleon.

\section{Nucleon emission functions}

\label{source}

The nucleon emission function describes the distributions of time, position,
and momentum at which nucleons are emitted. In the following, we first show
the distribution of nucleon emission times and the dependence of the average
emission time on nucleon momentum. This is then followed by the distribution
of nucleon emission positions and its dependence on nucleon momentum. The
momentum distribution of emitted nucleons is also given. All momenta are in
the center-of-mass system unless stated otherwise.

\subsection{Nucleon emission times}

The emission times of different particles in heavy ion collisions are
relevant for understanding both the collision dynamics and the mechanism for
particle production. In heavy ion collisions at intermediate energies, the
dynamics of nucleon emissions is mainly governed by the pressure of the
excited nuclear matter produced during the initial stage of collisions \cite%
{lis,pawel}. Since the stiff symmetry energy gives a larger pressure than
that due to the soft symmetry energy as shown in Fig. \ref{Psym}, it leads
to a faster emission of neutrons and protons. The relative emission sequence
of neutrons and protons is, however, determined by the difference in their
symmetry potentials. With the symmetry potential being generally repulsive
for neutrons and attractive for protons, neutrons are expected to be emitted
earlier than protons. The difference in neutron and proton emission times is
particularly large for the soft symmetry potential as its magnitude at
densities below the normal nuclear matter density, where most nucleons are
emitted, is larger than that of stiff symmetry potential as seen in Fig. \ref%
{SymPot}. Since the stiff symmetry potential changes sign when nuclear
density drops below $0.5\rho _{0}$, the relative emission sequence of
neutrons and protons in this case depends on the average density at which
they are emitted. Also, the emission time for protons is affected by the
repulsive Coulomb potential. Of course, details of these competing effects
on nucleon emissions depend on both the reaction dynamics and momenta of
nucleons. These details can only be studied by using transport models. 
\begin{figure}[th]
\includegraphics[scale=1.2]{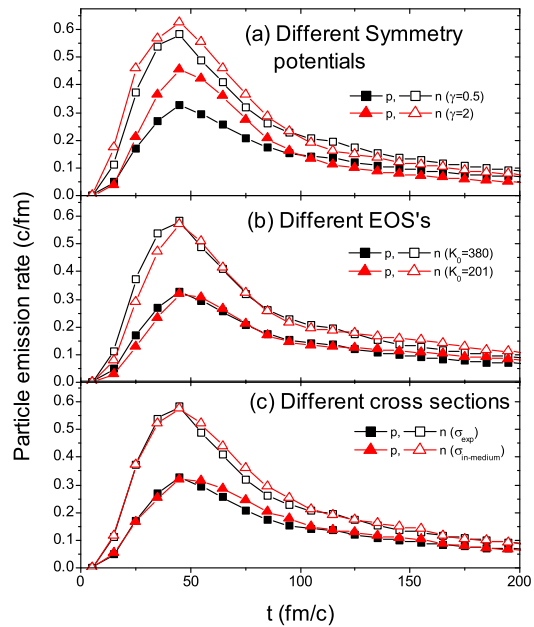} \vspace{0cm}
\caption{{\protect\small (Color online) Emission rates of protons and
neutrons as functions of time for different cases: (a) Soft or stiff
symmetry energy with }$K_{0}{\protect\small =380}${\protect\small \ MeV and }
$\protect\sigma _{\exp }${\protect\small ; (b) different EOS's, i.e., }$K_{0}%
{\protect\small =380}${\protect\small \ and }${\protect\small 200}$%
{\protect\small \ MeV with } $\protect\gamma =0.5${\protect\small ; and (c)
different \textsl{N}-\textsl{N} cross sections with }$K_{0}{\protect\small %
=380}${\protect\small \ MeV and }$\protect\gamma =0.5$.}
\label{emProb}
\end{figure}

As an example, we study here central collisions of $^{52}$Ca + $^{48}$Ca at $%
E=80$ \textrm{MeV/nucleon}. This particular reaction system with isospin
asymmetry $\delta =0.2$ can be studied at RIA. In the present work, nucleons
are considered as being emitted when their local densities are less than $%
\rho _{0}/8$ and subsequent interactions does not cause their recapture into
regions of higher density. Other emission criteria, such as taking the
nucleon emission time as its last collision time in the IBUU model, do not
change our conclusions. In Fig. \ref{emProb}, we show the emission rates of
protons and neutrons as functions of time for different cases. It is clearly
seen that there are two stages of nucleon emissions; an early fast emission
and a subsequent slow emission. This is consistent with the long-lived
nucleon emission source observed in previous BUU calculations \cite{handzy95}%
. In Fig. \ref{emProb} (a), effects due to different symmetry energies are
shown, and we find that the stiff symmetry energy (triangles) enhances the
emission of early high momentum protons and neutrons but suppresses late
slow emission compared with results from the soft symmetry energy (squares).
Difference between the emission rates of protons and neutrons is, however,
larger for the soft symmetry energy. Both features are what we have expected
from above discussions about the effects of nuclear symmetry energy on
particle emissions. Fig. \ref{emProb} (b) shows results by using different
incompressibilities of $K_{0}=380$ (squares) and $201$ (triangles) \textrm{%
MeV} for the isoscalar potential. It is seen that the nucleon emission rate
is not sensitive to the incompressibility $K_{0}$ of the symmetric nuclear
matter \textrm{EOS}, except that the stiff \textrm{EOS} slightly enhances
the nucleon emission rate at early stage of reactions. In Fig. \ref{emProb}
(c), we compare results from the experimental free space \textsl{N}-\textsl{N%
} cross section $\sigma _{\exp }$ (squares) and those from the in-medium 
\textsl{N}-\textsl{N} cross sections $\sigma _{\text{\textrm{in-medium}}}$
(triangles). Again, the nucleon emission rate is found to be insensitive to
the \textsl{N}-\textsl{N} cross sections used for the present reaction
system, except that the in-medium cross sections slightly enhance the
nucleon emission rate at later stage of reactions ($50\sim 100$ \textrm{fm/c}%
). 
\begin{figure}[th]
\includegraphics[scale=0.9]{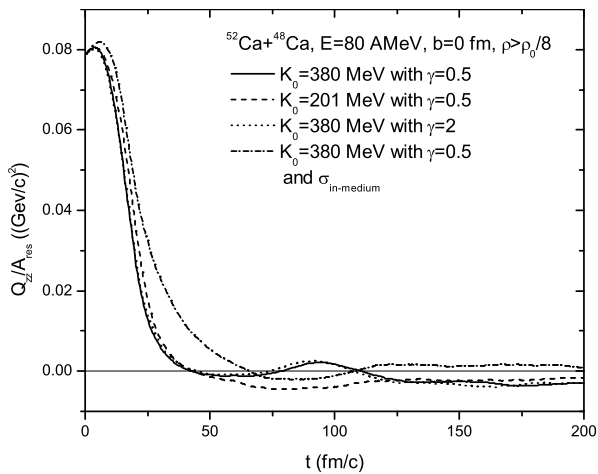} \vspace{0cm}
\caption{{\protect\small Time evolutions of scaled quadrupole moment of
nucleon momentum distribution} $Q_{zz}/A_{\mathrm{res}}${\protect\small \ by
using }$K_{0}=380${\protect\small \ and } $210${\protect\small \ \ MeV with }%
$\protect\gamma =0.5${\protect\small , } $K_{0}=380${\protect\small \ with }$%
\protect\gamma =2${\protect\small , and } $K_{0}=380${\protect\small \ MeV\
with }$\protect\gamma =0.5$ {\protect\small \ and in-medium \textsl{N}-%
\textsl{N} cross sections.}}
\label{Qzz}
\end{figure}

To see if nucleons are emitted from an equilibrium or non-equilibrium
sources, we consider time evolution of the scaled quadrupole moment of
nucleon momentum distribution $Q_{\mathrm{zz}}/A_{\mathrm{res}}$ defined as %
\cite{li95} 
\begin{equation}
Q_{\mathrm{zz}}/A_{\mathrm{res}}(t)=1/A_{res}\int \frac{d\mathbf{r}d\mathbf{p%
}}{(2\pi )^{3}}(2p_{z}^{2}-p_{x}^{2}-p_{y}^{2})f(\mathbf{r},\mathbf{p},t),
\label{QzzEq}
\end{equation}%
where $A_{\mathrm{res}}$ is the mass number of the residue composed of
nucleons at local density larger than $\rho _{0}/8$ and $f(\mathbf{r},%
\mathbf{p},t)$ is the nucleon phase-space distribution function given by the
IBUU model. How far the emission source deviates from thermal equilibrium is
then given by the value of $Q_{\mathrm{zz}}$, with $Q_{\mathrm{zz}}=0$
implying that thermal equilibrium is achieved. Shown in Fig. \ref{Qzz} are
time evolutions of $Q_{zz}/A_{res}$ for different nuclear symmetry
potentials, isoscalar potentials, and \textsl{N}-\textsl{N} cross sections.
It is seen that the emission source nearly reaches thermal equilibrium after
around $40$ \textrm{fm/c} except the case with in-medium \textsl{N}-\textsl{N%
} cross sections (dash-dotted curve). In the latter case, the thermal
equilibration time is delayed to about $70$ \textrm{fm/c} as a result of
smaller in-medium \textsl{N}-\textsl{N} cross sections than free \textsl{N}-%
\textsl{N} cross sections. Therefore, early emitted nucleons are mostly from
pre-equilibrium stage of heavy ion collisions, while late emitted ones are
more like statistical emissions. Since positive $Q_{\mathrm{zz}}$ indicates
incomplete nuclear stopping or nuclear transparency while negative one
implies transverse expansion or collectivity, all cases except the one with
in-medium \textsl{N}-\textsl{N} cross sections thus show strong nuclear
stopping but weak transverse expansion after $120$ \textrm{fm/c}. 
\begin{figure}[th]
\includegraphics[scale=0.9]{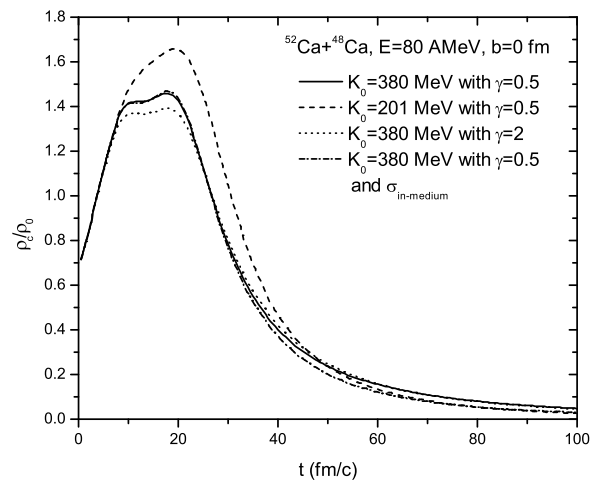} \vspace{0cm}
\caption{{\protect\small Time evolution of average central density by using} 
$K_{0}={\protect\small 380}${\protect\small \ and } ${\protect\small 210}$ 
{\protect\small \ MeV with }$\protect\gamma ={\protect\small 0.5}$%
{\protect\small , }$K_{0}=380${\protect\small \ with }$\protect\gamma =2,$ 
{\protect\small and }$K_{0}=380${\protect\small \ MeV\ with }$\protect\gamma %
=0.5${\protect\small \ and in-medium \textsl{N}-\textsl{N} cross sections.}}
\label{cdenTime}
\end{figure}

The weak dependence of the nucleon emission rate on the stiffness of
isoscalar potential is due to reduced difference in the pressure of excited
nuclear matter as a result of different maximum densities reached in
collisions, with the stiff one giving a lower density than the soft one.
This can be seen in Fig. \ref{cdenTime} where the time evolution of average
central density (calculated in a sphere located at the center of colliding
nuclear matter with radius of $3$ \textrm{fm}) is shown by using $K_{0}=380$
(solid curve) or $201$ (dashed curve) \textrm{MeV} but with same soft
symmetry energy and free nucleon-nucleon cross sections. Maximum average
central densities reached in the collisions are about $1.46\rho _{0}$ and $%
1.66\rho _{0}$, respectively, for $K_{0}=380$ and $201$ \textrm{MeV},
resulting in similar pressures of about $8$ and $7$ \textrm{MeV/fm}$^{3}$,
respectively. In addition, the soft \textrm{EOS} leads to a slower expansion
and a longer higher density stage than the stiff \textrm{EOS}. On the other
hand, the maximum average central density reached in the collisions is
reduced only slightly if the stiff symmetry energy is used for the case of $%
K_{0}=380$ MeV and free \textsl{N}-\textsl{N}cross sections (dotted curve).
Also shown in Fig. \ref{cdenTime} is the time evolution of average central
density by using $K_{0}=380$ \textrm{MeV} and the soft symmetry energy but
with in-medium \textsl{N}-\textsl{N} cross sections $\sigma _{\text{\textrm{%
in-medium}}}$ (dash-dotted curve). One finds that different \textsl{N}-%
\textsl{N} cross sections lead to almost similar time evolutions of average
central density. Although the free \textsl{N}-\textsl{N} cross sections,
which have larger values than the in-medium ones, would lead to more
stopping and thus higher density, enhanced Pauli blocking and mean-field
pressure effects bring the resulting maximum density similar to that
obtained using the in-medium cross sections. As a result, the two different
cross sections lead to similar pressures during the high density stage of
heavy ion collisions, and thus give comparable nucleon emission times as
shown in Fig. \ref{emProb} (c). 
\begin{figure}[th]
\includegraphics[scale=0.8]{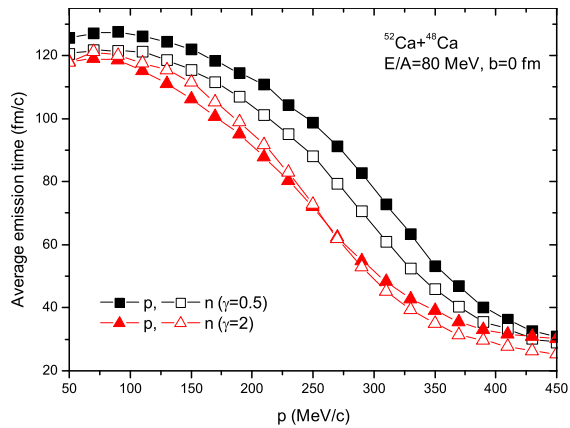} \vspace{0cm}
\caption{{\protect\small (Color online) Average emission times of protons
and neutrons as functions of their momenta for different symmetry energies.}}
\label{emTimeP}
\end{figure}

Energies of emitted nucleons are correlated to the times at which they are
emitted. Generally, earlier emitted nucleons have higher energies than later
emitted ones. This can be seen in Fig. \ref{emTimeP} where we show the
average emission times of protons and neutrons as functions of their momenta
by using $K_{0}=380$ \textrm{MeV} with the soft or stiff symmetry energy.
Indeed, high momentum nucleons are emitted in the early pre-equilibrium
stage of collisions when the average density is relatively high, while low
momentum ones are mainly emitted when the system is close to equilibrium and
the average density is low. Fig. \ref{emTimeP} further shows that the
average emission time of nucleons with a given momentum is earlier for the
stiff symmetry energy (triangles) than for the soft one (squares). Moreover,
there are significant delays in proton emissions (filled squares) with the
soft symmetry energy. These features are consistent with the results shown
in Fig. \ref{emProb}(a). For stiff symmetry energy, the change in the
emission sequence of neutrons and protons with their momenta is consistent
with the change in the sign of stiff symmetry potential at low densities as
shown in Fig. \ref{SymPot}. 
\begin{figure}[th]
\includegraphics[scale=1.2]{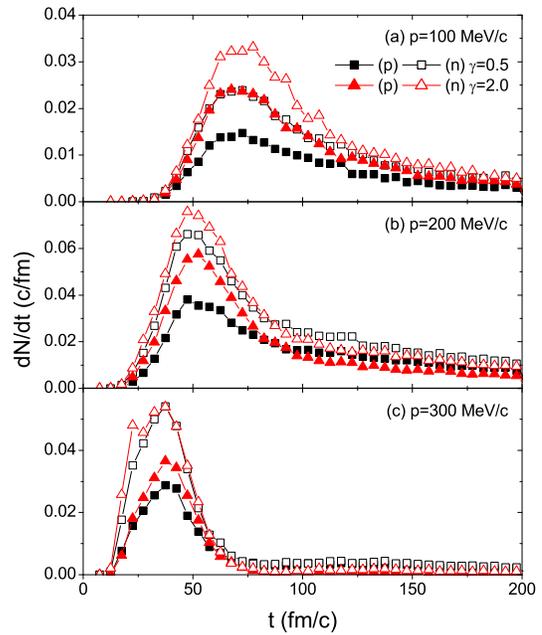} \vspace{0cm}
\caption{{\protect\small (Color online) Emission rates of neutrons and
protons with momenta 100 (a), 200 (b), and 300 (c) MeV/c (with bin size of
20 MeV/c) for $K_{0}=380$ MeV, free \textsl{N}-\textsl{N} cross sections,
and either soft or stiff symmetry energy.}}
\label{dNdt}
\end{figure}

We note, however, the distribution of the emission times for nucleons with
same momentum is quite broad particularly for low momentum nucleons. This is
shown in Fig. \ref{dNdt} for nucleons with momenta of $100$ (a), $200$ (b),
and $300$ (c) \textrm{MeV/c} (all with bin size of $20$ \textrm{MeV/c}),
obtained with $K_{0}=380$ MeV, free \textsl{N}-\textsl{N} cross sections,
and either soft or stiff symmetry energy. Although the peak emission time
increases with decreasing nucleon momentum, the distribution of nucleon
emission times is broad. Furthermore, at later times more nucleons with
momentum of $200$ \textrm{MeV/c} are emitted than those with momentum of $%
100 $ \textrm{MeV/c} are emitted. For both protons and neutrons, more are
emitted for the stiff than for the soft symmetry energy. 
\begin{figure}[th]
\includegraphics[scale=0.8]{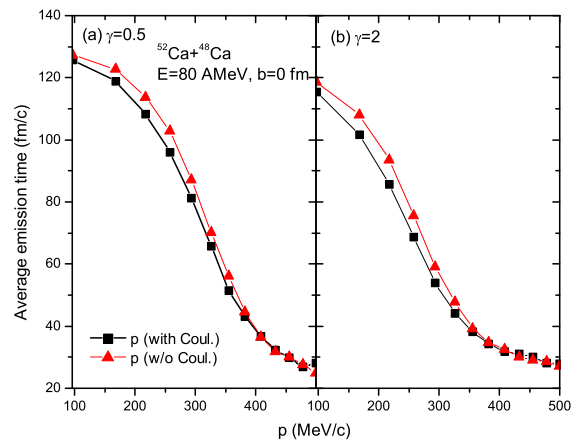} \vspace{0cm}
\caption{{\protect\small (Color online) Average emission time of protons as
a function of their momenta with (filled squares) and without (open squares)
Coulomb interaction by using the soft (a) or stiff (b) symmetry energy.}}
\label{NoCou}
\end{figure}

In low and intermediate energy heavy-ion collisions induced by heavy nuclei,
the Coulomb interaction is expected to affect the emission time of protons
and thus also the neutron-proton correlation function \cite{coul93,coul95}.
In present IBUU simulations, the Coulomb potential is included explicitly in
the dynamic evolution of colliding nuclei. Effects of Coulomb interaction on
the proton emission time can be seen in Fig. \ref{NoCou}, where the average
emission time of protons is shown as a function of their momenta with
(filled squares) and without (open squares) Coulomb interaction by using
either the soft (left panel) or stiff (right panel) symmetry energy. One
finds that the Coulomb interaction is not important for $^{52}$Ca + $^{48} $%
Ca as it only shortens the proton average emission time by factors less than 
$10\%$ for both the soft and stiff symmetry energies. The Coulomb potential
effect is even smaller for protons with high momentum, which are emitted
early during collisions. We note that the Coulomb potential does not affect
the neutron emission time as expected.

\subsection{Size of nucleon emission source}

\begin{figure}[th]
\includegraphics[scale=0.8]{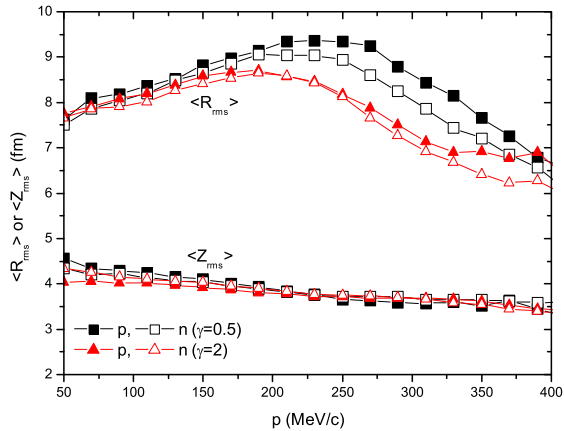} \vspace{0cm}
\caption{{\protect\small (Color online) Root-mean-squared transverse
emission radius} $\langle R_{\mathrm{rms}}\rangle $ {\protect\small \ and
longitudinal emission radius} $\langle Z_{\mathrm{rms}}\rangle $%
{\protect\small \ of neutrons and protons as functions of their momenta by
using }$K_{0}=380$ {\protect\small \ MeV with either soft (squares) or stiff
(triangles) symmetry energy.}}
\label{RxyzP}
\end{figure}

The size of emission source can be characterized by the root-mean-squared
radii of emitted protons and neutrons in the transverse ($\langle R_{\mathrm{%
rms}}\rangle $) and longitudinal ($\langle Z_{\mathrm{rms}}\rangle $)
directions. In Fig. \ref{RxyzP}, we show these radii as functions of nucleon
momentum using $K_{0}=380$ MeV and free \textsl{N}-\textsl{N} cross sections
but with either soft (squares) or stiff (triangles) symmetry energy. It is
seen that for both symmetry energies the emission source has an oblate shape
with a larger transverse radius than longitudinal radius. Furthermore, the
emission source size is larger for the soft than the stiff symmetry energy.
However, the difference between the transverse radii of proton and neutron
emission sources is more appreciable in the case of soft symmetry energy
than stiff symmetry energy. These features are consistent with the emission
times shown in Fig. \ref{emTimeP}. 
\begin{figure}[th]
\includegraphics[scale=1.2]{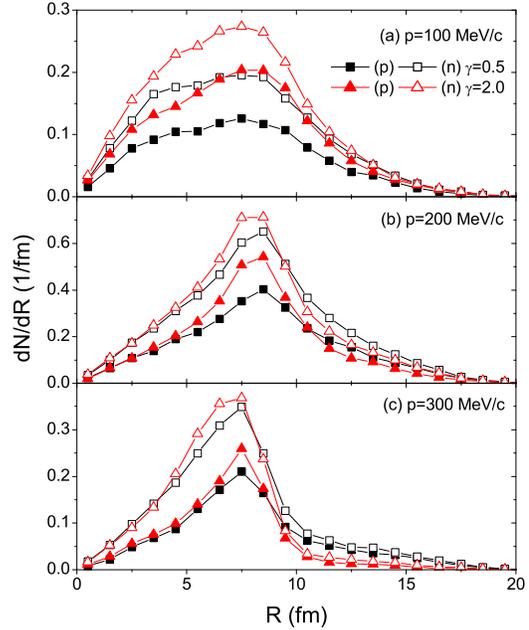} \vspace{0cm}
\caption{{\protect\small (Color online) Transverse radius distribution of
emitted nucleons with momenta 100 (a), 200 (b), and 300 (c) MeV/c (with bin
size of 20 MeV/c) for $K_{0}=380$ MeV, free \textsl{N}-\textsl{N} cross
sections, and either soft or stiff symmetry energy.}}
\label{dNdR}
\end{figure}

The decrease of the transverse radius with increasing nucleon momentum shown
in Fig. \ref{RxyzP} is expected as higher momentum nucleons are emitted
earlier when the source size is more compact. The small transverse radius of
the emission source for lower momentum nucleons are, on the other hand, due
to the broad distributions in nucleon emission times, resulting in the
emissions of more intermediate momentum nucleons than low momentum nucleons
at later times. As a result, the source size for low momentum nucleons is
largely determined by earlier emitted nucleons. This can also been seen in
Fig. \ref{dNdR} where we show the transverse radius distributions of emitted
nucleons with momenta $100$ (a), $200$ (b), and $300$ (c) \textrm{MeV/c}
(with bin size of $20$ \textrm{MeV/c}) for $K_{0}=380$ MeV, free \textsl{N}-%
\textsl{N} cross sections, and either soft or stiff symmetry energy. It is
seen that the peak of the transverse radius distribution first increases
then decreases as the nucleon momentum gets larger, consistent with that
shown in Fig. \ref{RxyzP}. Furthermore, width of the radius distribution is
narrower for high momentum nucleons than for low momentum ones. 
\begin{figure}[th]
\includegraphics[scale=1.0]{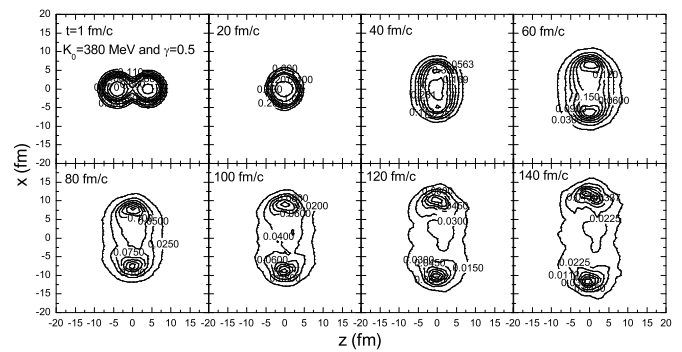} \vspace{0cm}
\caption{{\protect\small Density contours }$\protect\rho (x,0,z)$ 
{\protect\small \ at different times by using }$K_{0}=380$ {\protect\small \
MeV and free N-N cross sections with the soft symmetry energy.}}
\label{DenXZ}
\end{figure}
\begin{figure}[th]
\includegraphics[scale=1.0]{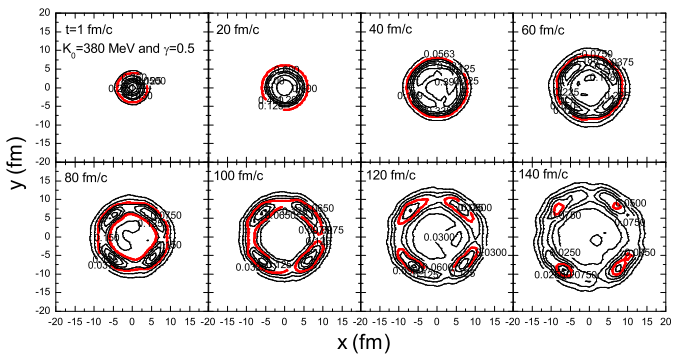} \vspace{0cm}
\caption{{\protect\small (Color online) Density contours }$\protect\rho %
(x,y,0)$ {\protect\small \ at different times by using }$K_{0}=380$ 
{\protect\small \ MeV and free N-N cross sections with the soft symmetry
energy. The thick curves represent }$\protect\rho _{0}/8$.}
\label{DenXY}
\end{figure}
\begin{figure}[th]
\includegraphics[scale=0.8]{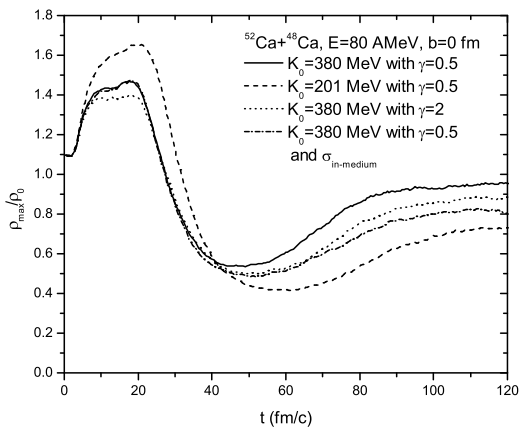} \vspace{0cm}
\caption{{\protect\small Time evolution of maximum density for different
incompressibilities, \textsl{N}-\textsl{N} cross sections, and symmetry
energies as in Fig. \ref{cdenTime}}}
\label{MaxDen}
\end{figure}

To see more clearly the size and structure of nucleon emission source, we
show in Figs. \ref{DenXZ} and \ref{DenXY} the density contours $\rho (x,0,z)$
in the reaction plane and $\rho (x,y,0)$ perpendicular to the reaction plane
at different times by using $K_{0}=380$ \textrm{MeV} and free \textsl{N}-%
\textsl{N} cross sections with the soft symmetry energy. It is seen that
after initial contact of target and project at $1$ \textrm{fm/c}, the
nuclear matter is compressed to high density during the first $20$ \textrm{%
fm/c}. This is followed by expansions, mainly in the transverse direction,
leading to the formation of an oblate bubble nuclear matter, which
subsequently changes into a ring-shaped structure. The ring structure has a
long lifetime and decays into several fragments after about $100$ \textrm{%
fm/c}. Such exotic structures also exist in the other cases, i.e., using $%
K_{0}=201$ \textrm{MeV}, stiff symmetry energy, or in-medium \textsl{N}-%
\textsl{N} cross sections. This interesting phenomenon is not new and was
studied extensively a decade ago. For medium mass nuclei, this is a common
phenomenon in central or near central collisions at energy we consider here %
\cite{Moretto92,Bauer92,handzy95c}. The formation of bubble and ring
structures during expansion implies that the central density shown in Fig. %
\ref{cdenTime} is not always the maximum density reached in collisions. In
Fig. \ref{MaxDen}, we show the time evolution of maximum density for
different incompressibilities, \textsl{N}-\textsl{N} cross sections, and
symmetry energies as in Fig. \ref{cdenTime}. We see that after about $30$ 
\textrm{fm/c} when the bubble is formed the maximum density has a value
around normal nuclear density and is very different from the central density
shown in Fig. \ref{cdenTime}. Since low momentum nucleons are mainly emitted
during the late stage of heavy ion collisions from the ring and its decaying
fragments, the symmetry energy effect is thus small as a result of the small
difference between soft and stiff symmetry potentials at normal nuclear
density as shown in Fig. \ref{SymPot}.

\subsection{Momentum distributions of emitted nucleons}

\begin{figure}[th]
\includegraphics[scale=0.8]{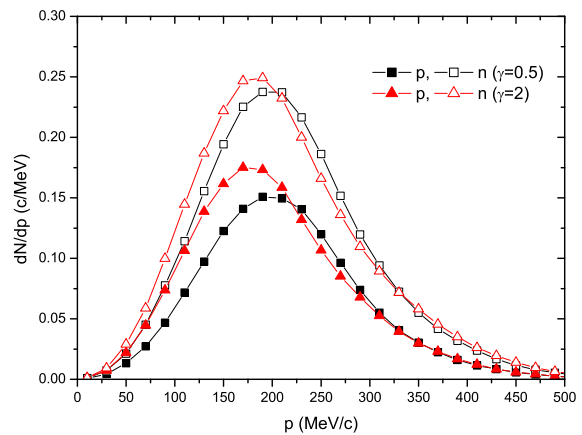} \vspace{0cm}
\caption{{\protect\small (Color online) Momentum distributions of emitted
neutrons and protons for }$K_{0}=380${\protect\small \ MeV and free \textsl{N%
}-\textsl{N} cross sections with soft (squares) or stiff (triangles)
symmetry energy.}}
\label{XNp}
\end{figure}

Fig. \ref{XNp} shows the momentum distributions of emitted neutrons and
protons for $K_{0}=380$ \textrm{MeV} and free \textsl{N}-\textsl{N} cross
sections with soft (squares) or stiff (triangles) symmetry energy. It is
seen that most nucleons are emitted with momenta around $p=180$ \textrm{MeV/c%
}, and the peak momentum shifts to a larger value as the symmetry energy
becomes softer. The lower peak momentum in the case of stiff symmetry energy
is due to the larger potential energy during initial compressions and thus
lower average nucleon kinetic energy as a result of energy conservation.
Furthermore, symmetry energy effects on lower momentum nucleons are stronger
for protons than for neutrons, which is consistent with the larger variation
of proton symmetry potential at low densities with the stiffness of symmetry
energy than that of neutron symmetry potential as shown in Fig. \ref{SymPot}%
. 

\section{Nucleon-nucleon correlation functions}

\label{correlation}

The nucleon emission function, which is important for understanding the
reaction dynamics in heavy ion collisions, can be extracted from
two-particle correlation functions; see, e.g., Refs. \cite%
{Boal90,bauer,ardo97,Wied99} for reviews. In most studies, only the
two-proton correlation function has been measured \cite%
{gong90,gong91,gong93,kunde93,handzy95}. Recently, data on two-neutron and
neutron-proton correlation functions have also become available. The
neutron-proton correlation function is especially useful as it is free of
correlations due to wave-function anti-symmetrization and Coulomb
interactions. Indeed, Ghetti \textit{et al.} have deduced from measured
neutron-proton correlation function the time sequence of neutron and proton
emissions \cite{Ghetti00,Ghetti01}.

In standard Koonin-Pratt formalism \cite{koonin77,pratt1,pratt2}, the
two-particle correlation function is obtained by convoluting the emission
function $g(\mathbf{p},x)$, i.e., the probability for emitting a particle
with momentum $\mathbf{p}$ from the space-time point $x=(\mathbf{r},t)$,
with the relative wave function of the two particles, i.e., 
\begin{equation}
C(\mathbf{P},\mathbf{q})=\frac{\int d^{4}x_{1}d^{4}x_{2}g(\mathbf{P}%
/2,x_{1})g(\mathbf{P}/2,x_{2})\left| \phi (\mathbf{q},\mathbf{r})\right|^{2}%
}{\int d^{4}x_{1}g(\mathbf{P}/2,x_{1})\int d^{4}x_{2}g(\mathbf{P}/2,x_{2})}.
\label{Eq1}
\end{equation}%
In the above, $\mathbf{P(=\mathbf{p}_{1}+\mathbf{p}_{2})}$ and $\mathbf{q(=}%
\frac{1}{2}(\mathbf{\mathbf{p}_{1}-\mathbf{p}_{2}))}$ are, respectively, the
total and relative momenta of the particle pair; and $\phi (\mathbf{q},%
\mathbf{r})$ is the relative two-particle wave function with $\mathbf{r}$
being their relative position, i.e., $\mathbf{r=(r}_{2}\mathbf{-r}_{1}%
\mathbf{)-}$ $\frac{1}{2}(\mathbf{\mathbf{v}_{1}+\mathbf{v}_{2})(}t_{2}-t_{1}%
\mathbf{)}$. This approach has been very useful in studying effects of
nuclear equation of state and nucleon-nucleon cross sections on the reaction
dynamics of intermediate energy heavy-ion collisions \cite{bauer}. In the
present paper, we use the Koonin-Pratt method to determine the
nucleon-nucleon correlation functions in order to study the effect due to
the density dependence of nuclear symmetry energy on the spatial and
temporal structure of nucleon emission source in intermediate energy heavy
ion collisions.

\subsection{Symmetry energy effects}

\begin{figure}[th]
\includegraphics[scale=1.1]{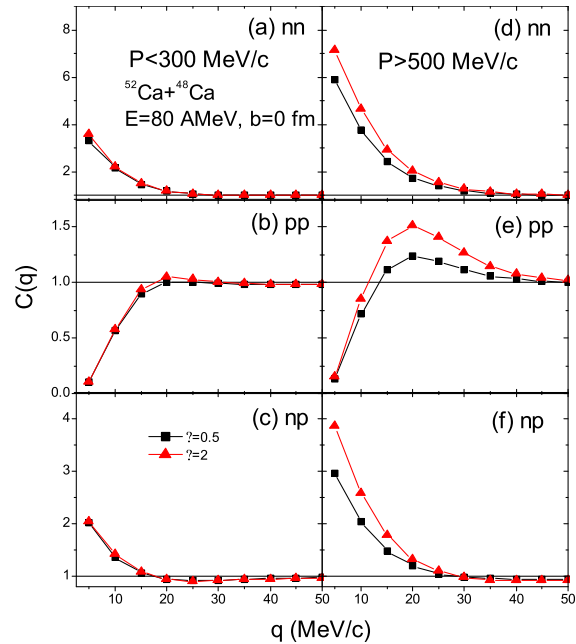} \vspace{0cm}
\caption{{\protect\small (Color online) Two-nucleon correlation functions
gated on total momentum of nucleon pairs using the soft (filled squares) or
stiff (filled triangles) symmetry energy. Left panels are for $P<300$ MeV/c
while right panels are for $P>500$ MeV/c.}}
\label{CFsym}
\end{figure}

Using the program Correlation After Burner \cite{hbt}, which takes into
account final-state nucleon-nucleon interactions, we have evaluated
two-nucleon correlation functions from the emission function given by the
IBUU model. Shown in Fig. \ref{CFsym} are two-nucleon correlation functions
gated on total momentum $P$ of nucleon pairs from central collisions of $%
^{52}$Ca + $^{48}$Ca at $E=80$ \textrm{MeV/nucleon}. The left and right
panels are for $P<300$ \textrm{MeV/c} and $P>500$ \textrm{MeV/c},
respectively. For both neutron-neutron (upper panels) and neutron-proton
(lower panels) correlation functions, they peak at $q\approx 0$ \textrm{MeV/c%
}. The proton-proton correlation function (middle panel) is, however, peaked
at about $q=20$ \textrm{MeV/c} due to strong final-state s-wave attraction,
but is suppressed at $q=0$ as a result of Coulomb repulsion and
anti-symmetrization of two-proton wave function. These general features are
consistent with those observed in experimental data from heavy ion
collisions \cite{Ghetti00}. 

Since emission times of low-momentum nucleons do not change much with the
different $E_{\mathrm{sym}}(\rho )$ used in IBUU model as shown in Fig. \ref%
{emTimeP}, two-nucleon correlation functions are not much affected by the
stiffness of symmetry energy. On the other hand, the emission times of
high-momentum nucleons, which are dominated by those with momenta near $250$ 
\textrm{MeV/c}, differ appreciably for the two symmetry energies considered
here. Correlation functions of high-momentum nucleon pairs thus show a
strong dependence on nuclear symmetry energy. Gating on nucleon pairs with
high total momentum allows one to select those nucleons that have short
average spatial separations at emissions and thus exhibit enhanced
correlations. For these nucleon pairs with high total momentum, strength of
their correlation function is stronger for the stiff symmetry energy than
for the soft symmetry energy: about $30\%$ and $20\%$ for neutron-proton
pairs and neutron-neutron pairs at low relative momentum $q=5$ MeV/c,
respectively, and $20\%$ for proton-proton pairs at $q=20$ \textrm{MeV/c}.
The neutron-proton correlation function thus exhibits the highest
sensitivity to variations of $E_{\mathrm{sym}}(\rho )$. As shown in Fig. \ref%
{emTimeP} and discussed earlier, the emission sequence of neutrons and
protons is sensitive to $E_{\mathrm{sym}}(\rho )$. With stiff $E_{\mathrm{sym%
}}(\rho )$, high momentum neutrons and protons are emitted almost
simultaneously, and they are thus temporally strongly correlated, leading to
a larger neutron-proton correlation function. On the other hand, proton
emissions in the case of soft $E_{\mathrm{sym}}(\rho )$ are delayed compared
to neutrons, so they are temporally weakly correlated with neutrons. The
resulting values of neutron-proton correlation function are thus smaller.
Furthermore, both neutrons and protons are emitted earlier with stiff $E_{%
\mathrm{sym}}(\rho )$, so they have shorter spatial separation at emissions.
As a result, neutrons and protons are more correlated for the stiff symmetry
energy than for the soft symmetry energy. Our results thus clearly
demonstrate that correlation functions of nucleon pairs with high total
momentum can indeed reveal sensitively the effect of nuclear symmetry energy
on the temporal and spatial distributions of emitted nucleons. 
\begin{figure}[th]
\includegraphics[scale=1.1]{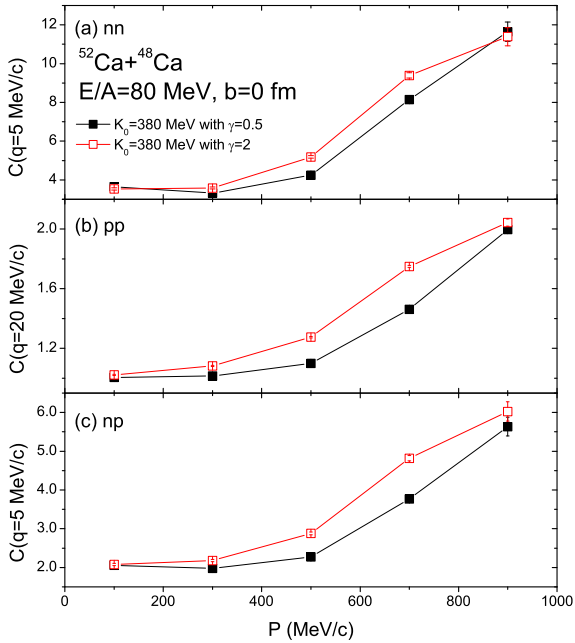} \vspace{0cm}
\caption{{\protect\small (Color online) Total momentum dependence of the
correlation functions for neutron-neutron (a) and neutron-proton (c) pairs
with relative momentum }$q=5${\protect\small \ MeV/c and proton-proton (b)
pairs with relative momentum }$q=20${\protect\small \ MeV/c by using }$%
K_{0}=380${\protect\small \ MeV with both soft (filled squares) and stiff
(open squares) symmetry energies.}}
\label{PtotDep}
\end{figure}

To see more clearly the symmetry energy effect on nucleon-nucleon
correlation functions, we show in Fig. \ref{PtotDep} the total momentum
dependence of the correlation functions for neutron-neutron and
neutron-proton pairs with relative momentum $q=5$ \textrm{MeV/c} and
proton-proton pairs with relative momentum $q=20$ \textrm{MeV/c} by using $%
K_{0}=380$ \textrm{MeV} with both soft (filled squares) and stiff (open
squares) symmetry energies. It is seen that the symmetry energy effect is
more pronounced for pairs with higher total momentum as discussed
previously. However, the effect becomes weaker when their total momentum
becomes very large. This is due to the fact that very energetic neutrons and
protons mainly come from very early stage of collisions when the reaction
system reaches maximum compression and nucleon emissions are largely
affected by the isospin-independent part of \textrm{EOS}. The symmetry
energy effect on the emission time of nucleons is therefore reduced for
these very energetic neutrons and protons as shown in Fig. \ref{emTimeP}.
Furthermore, emission times of very energetic neutrons and protons are
strongly influenced by the high density behavior of symmetry energy, and for
the stiff symmetry energy very high momentum neutrons are emitted earlier
than protons as shown in Fig. \ref{emTimeP}. As a result, the symmetry
energy effect on the correlation function of very energetic neutron-proton
pairs is also reduced. 
\begin{figure}[th]
\includegraphics[scale=1.1]{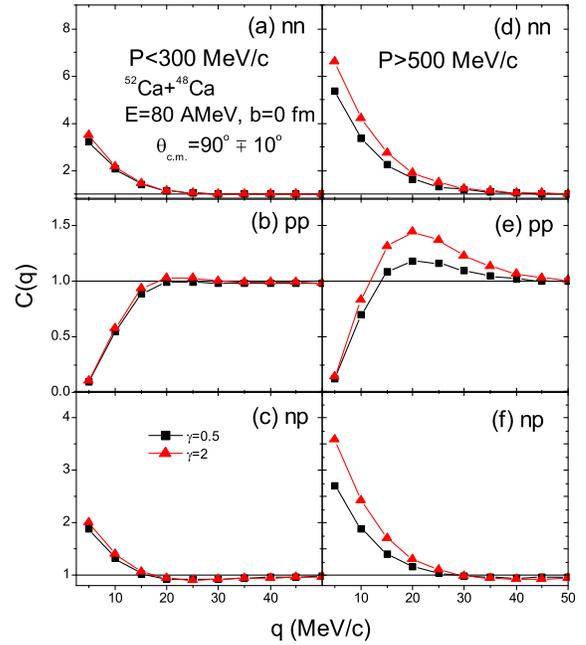} \vspace{0cm}
\caption{{\protect\small (Color online) Same as Fig. \ref{CFsym} but with
angular cut }$\protect\theta _{\text{\textrm{c.m.}}}=90^{\circ }\pm
10^{\circ }$ {\protect\small in center-of-mass system.}}
\label{CFTrans}
\end{figure}
\begin{figure}[th]
\includegraphics[scale=1.1]{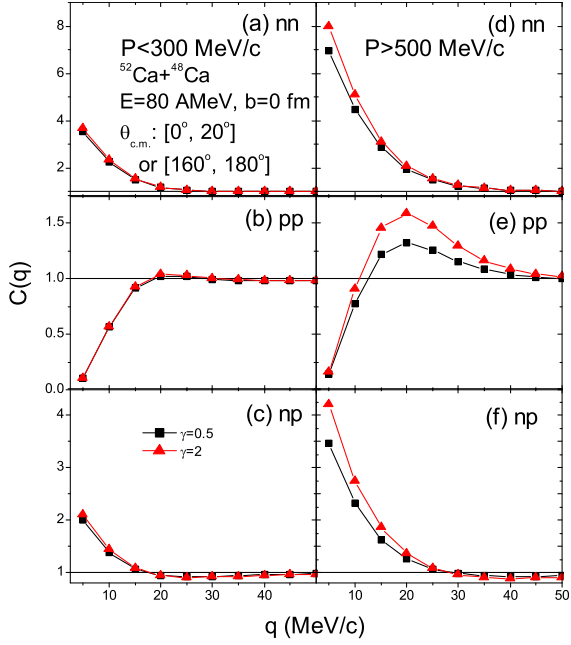} \vspace{0cm}
\caption{{\protect\small (Color online) Same as Fig. \ref{CFsym} but with
angular cut }$\protect\theta _{\text{\textrm{c.m.}}}=0^{\circ }\sim
20^{\circ }$ {\protect\small or }$160^{\circ }\sim 180^{\circ }$ 
{\protect\small in center-of-mass system.}}
\label{CFLong}
\end{figure}

Since the nuclear symmetry energy affects differently the transverse and
longitudinal radii of nucleon emission source, it is useful to study also
the dependence of two-nucleon correlation functions on the direction of
emitted nucleons. We have thus considered both transverse cut with $\theta _{%
\text{\textrm{c.m.}}}=90^{\circ }\pm 10^{\circ }$ and longitudinal cut with $%
\theta _{\text{\textrm{c.m.}}}=0^{\circ }\sim 20^{\circ }$ or $160^{\circ
}\sim 180^{\circ }$, where $\theta _{\mathrm{c.m.}}$ is the center-of-mass
angle with respect to the beam direction. Results similar to Fig. \ref{CFsym}
for two-proton, two-neutron, and neutron-proton correlation functions using
both stiff and soft symmetry energies but stiff isospin independent EOS and
free \textsl{N}-\textsl{N} cross sections are shown in Fig. \ref{CFTrans}
and Fig. \ref{CFLong} for transverse and longitudinal cuts, respectively. It
is seen that the transverse cut gives smaller values for nucleon-nucleon
correlation functions than the longitudinal cut. This is easy to understand
as the correlation functions with transverse (longitudinal) cut mainly
reflect the transverse (longitudinal) size or lifetime of emission source. A
transverse radius larger than the longitudinal radius shown in Fig. \ref%
{RxyzP} thus leads to smaller values for the nucleon-nucleon correlation
functions. Our results further show that the correlation functions of
nucleon pairs with total momentum $P>500$ MeV/c obtained with transverse cut
have stronger symmetry energy dependence than those obtained with
longitudinal cut. For transverse cut, the symmetry energy effects are about $%
23\%$ and $33\%$ for neutron-neutron and neutron-proton pairs at $q=5$ 
\textrm{MeV/c}, and $22\%$ for proton-proton pairs at $q=20$ \textrm{MeV/c},
respectively, but corresponding effects are $14\%$, $22\%$, and $19\%$,
respectively, for longitudinal cut. However, compared to results in Fig. \ref%
{CFsym} without the angular cut, there is only a slight enhancement of
two-nucleon correlation functions when the transverse cut is imposed, as
most nucleons are emitted in the transverse direction. 
\begin{figure}[th]
\includegraphics[scale=1.1]{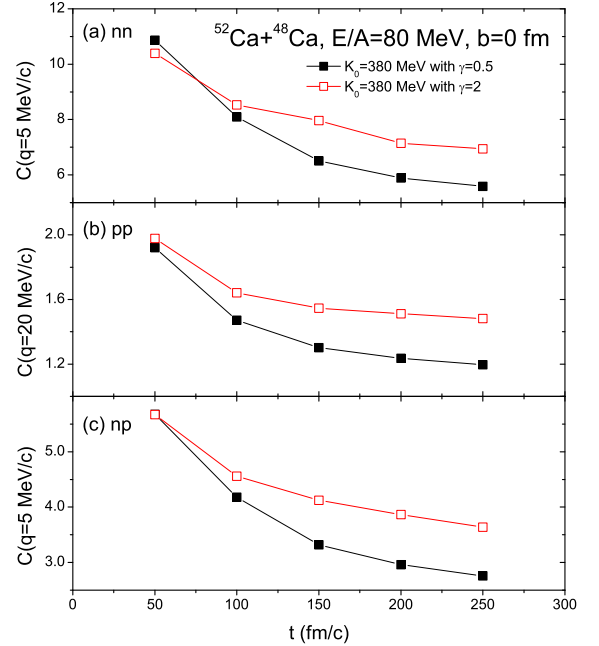} \vspace{0cm}
\caption{{\protect\small (Color online) Dependence of two-nucleon
correlation functions on stop times for high total momentum ($P>500$ \textrm{%
MeV}) neutron-neutron (a) and neutron-proton (c) pairs with relative
momentum }$q=5${\protect\small \ MeV/c and proton-proton (b) pairs with
relative momentum }$q=20${\protect\small \ MeV/c by using } $K_{0}=380$%
{\protect\small \ MeV with soft (filled squares) or stiff (open squares)
symmetry energy.}}
\label{CFtime}
\end{figure}

In the above analysis, the IBUU calculations are terminated at $200$ \textrm{%
fm/c}. In Fig. \ref{CFtime}, we show the stop time dependence of the
correlation functions for high total momentum ($P>500$ MeV) neutron-neutron
and neutron-proton pairs with relative momentum $q=5$ \textrm{MeV/c }and
proton-proton pairs with relative momentum $q=20$ \textrm{MeV/c} by using $%
K_{0}=380$ \textrm{MeV} with soft (filled squares) or stiff (open squares)
symmetry energy. It is seen that values of these correlation functions are
essentially fixed after about $150$ \textrm{fm/c} for both soft (open
squares) and stiff (filled squares) symmetry energies. Furthermore, the
symmetry energy effect starts to appear after $50$ \textrm{fm/c}, when the
central density is about $1/5\rho_0$ of normal nuclear matter density as
shown in Fig. \ref{cdenTime} and the maximum density is about $1/2\rho_0$ as
shown in Fig. \ref{MaxDen}, indicating that two-nucleon correlation
functions are sensitive to the properties of nuclear symmetry energy at
densities around and below the normal nuclear density.

\subsection{EOS and \textsl{N}-\textsl{N} cross section effects}

\begin{figure}[th]
\includegraphics[scale=1.1]{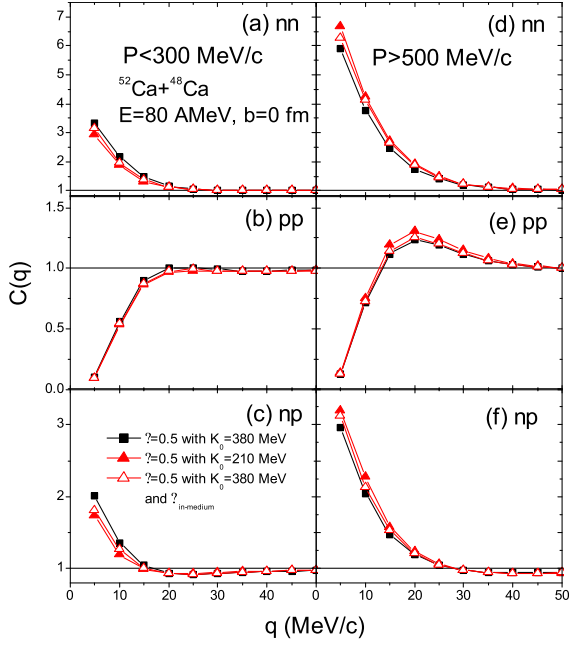} \vspace{0cm}
\caption{{\protect\small (Color online) Two-nucleon correlation functions
gated on total momentum of nucleon pairs using the soft symmetry energy but
different values of incompressibility }$K_{0}${\protect\small \ and \textsl{N%
}-\textsl{N} cross sections. Left panels are for $P<300$ MeV/c while right
panels are for $P>500$ MeV/c.}}
\label{CFeoscrsc}
\end{figure}

We have also studied the effects of isoscalar potential and in-medium 
\textsl{N}-\textsl{N} cross sections on nucleon-nucleon correlation
functions. In Fig. \ref{CFeoscrsc}, we show gated two-nucleon correlation
functions from central collisions of $^{52}$Ca + $^{48}$Ca at $E=80$ \textrm{%
MeV/nucleon} using the soft symmetry energy but different values of
incompressibility $K_{0}$ and \textsl{N}-\textsl{N} cross sections. One
finds that reducing $K_{0}$ from $380$ (filled squares) to $201$ (filled
triangles) \textrm{MeV} increases the value of the correlation function for
high momentum pairs ($P>500$ \textrm{MeV/c}) but decreases that for low
momentum pairs ($P<300$ \textrm{MeV/c}). Specifically, the increases are
about $8\%$ and $13\%$ for high momentum neutron-proton and neutron-neutron
pairs with relative momentum $q=5$ \textrm{MeV/c}, and about $5\%$ for high
momentum proton-proton pairs with relative momentum $q=20$ \textrm{MeV/c},
while corresponding decreases for low momentum pairs are about $14\%$, $3\%$%
, and $12\%$. Changing experimental free space \textsl{N}-\textsl{N} cross
section $\sigma _{\exp }$ \ to in-medium \textsl{N}-\textsl{N} cross
sections $\sigma _{\text{\textrm{in-medium}}}$ (open triangles) leads to
changes of about $5\%$, $6\%$, and $1\%$ in the neutron-proton,
neutron-neutron, and proton-proton correlation functions for high momentum
particle pairs ($P>500$ \textrm{MeV/c}), but about $10\%$, $2\%$, and $5\%$
for low momentum particle pairs ($P<300$ \textrm{MeV/c}). These results are
similar to those found in Ref. \cite{bauer}. The weak dependence of
two-nucleon correlation functions on the stiffness of isoscalar potential
and the \textsl{N}-\textsl{N} cross sections is consistent with the nucleon
emission times shown in the last section.

\subsection{Impact parameter and incident energy dependence}

\begin{figure}[th]
\includegraphics[scale=1.1]{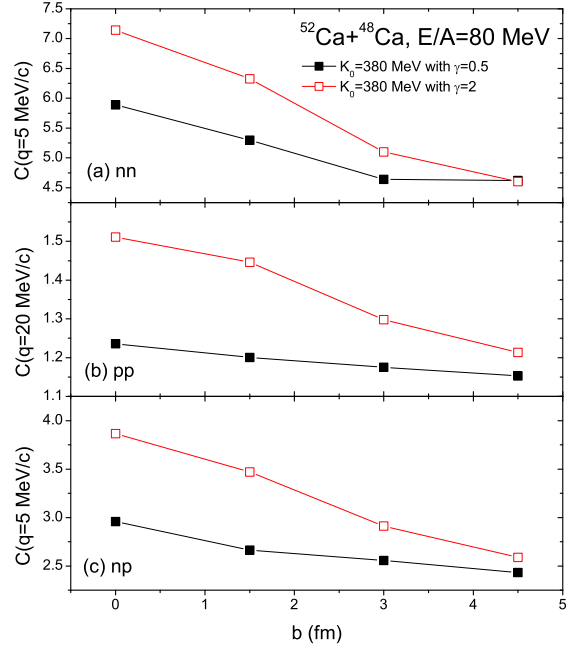} \vspace{0cm}
\caption{{\protect\small (Color online) Impact parameter dependence of
neutron-neutron (a) and neutron-proton (c) correlation functions at relative
momentum }$q=5${\protect\small \ MeV/c and proton-proton (b) correlation
function at relative momentum }$q=20${\protect\small \ MeV/c for particle
pairs with high total momentum }$P>500${\protect\small \ MeV/c by using }$%
K_{0}=380${\protect\small \ MeV and free \textsl{N}-\textsl{N} cross
sections with the soft (filled squares) or stiff (open squares) symmetry
energy.}}
\label{Bdep}
\end{figure}

The nucleon-nucleon correlation functions also depend on the impact
parameter and incident energy of heavy ion collisions. Fig. \ref{Bdep} shows
the impact parameter dependence of neutron-neutron and neutron-proton
correlation functions at relative momentum $q=5$ \textrm{MeV/c} and
proton-proton correlation function at relative momentum $q=20$ \textrm{MeV/c}
for particle pairs with high total momentum $P>500$ \textrm{MeV/c} by using $%
K_{0}=380$ \textrm{MeV} and free \textsl{N}-\textsl{N} cross sections with
the soft (filled squares) or stiff (open squares) symmetry energy. It is
seen that the symmetry energy effect is stronger in central and semi-central
collisions and becomes weaker in semi-peripheral and peripheral collisions.
This simply reflects the fact that nuclear compression is weaker in
peripheral collisions, and the symmetry energy effect mainly comes from the
small difference between the soft and stiff symmetry energies at low
densities. Also, one finds that values of these correlation functions
decrease with increasing impact parameter, which is consistent with previous
BUU results for the proton-proton correlation function \cite{gong91}. The
correlation function obtained with the stiff symmetry energy exhibits,
however, a stronger dependence on impact parameter than that using the soft
symmetry energy as the former has a stronger density dependence. 
\begin{figure}[th]
\includegraphics[scale=1.1]{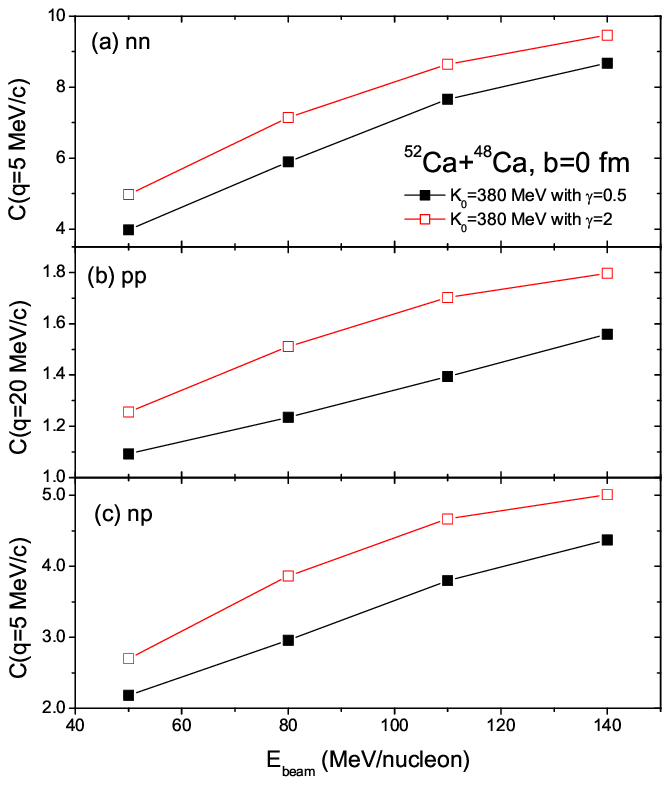} \vspace{0cm}
\caption{{\protect\small (Color online) Dependence of nucleon-nucleon
correlation functions on the incident energy for hight total momentum ($%
P>500 $ MeV) neutron-neutron (a) and neutron-proton (c) pairs with relative
momentum }$q=5${\protect\small \ MeV/c and proton-proton (b) pairs with
relative momentum }$q=20${\protect\small \ MeV/c by using }$K_{0}=380$%
{\protect\small \ MeV and free \textsl{N}-\textsl{N} cross sections with the
soft (filled squares) or stiff (open circles) symmetry energy.}}
\label{Edep}
\end{figure}

Dependence of nucleon-nucleon correlation functions on the incident energy
of heavy ion collisions is shown in Fig. \ref{Edep} for high total momentum (%
$P>500$ MeV) neutron-neutron and neutron-proton pairs with relative momentum 
$q=5$ \textrm{MeV/c }and for proton-proton pairs with relative momentum $q=20
$ \textrm{MeV/c}. The results are obtained using $K_{0}=380$ \textrm{MeV}
and free \textsl{N}-\textsl{N} cross sections with the soft (filled squares)
or stiff (open squares) symmetry energy. It is seen that the values of
nucleon-nucleon correlation functions increase with increasing incident
energy. This is understandable since nuclear compression increases with
increasing incident energy, and nucleons are also emitted earlier, leading
to a smaller source size. Also, the stiff symmetry energy gives a larger
value for the nucleon-nucleon correlation function than the soft symmetry
energy, and the difference does not depend much on incident energies,
although the relative effect is reduced. This is due to the fact that at
high incident energies the symmetry energy effect mainly comes from early
compression stage of collisions when nuclear pressure affects strongly the
emission times of nucleons. Furthermore, the system expands rapidly at high
incident energies, the low density behavior of symmetry potential thus plays
a less important role on the relative emission time of neutrons and protons.
According to Fig. \ref{emTimeP}, nucleons are emitted earlier with stiff
rather than with soft symmetry energies. Therefore, the symmetry potential
effects remain appreciable with increasing incident energy. On the other
hand, the initial compression is relatively weak at lower incident energies,
and the behavior of symmetry energy at low densities becomes relevant.

\subsection{Reaction system dependence}

\begin{figure}[th]
\includegraphics[scale=1.1]{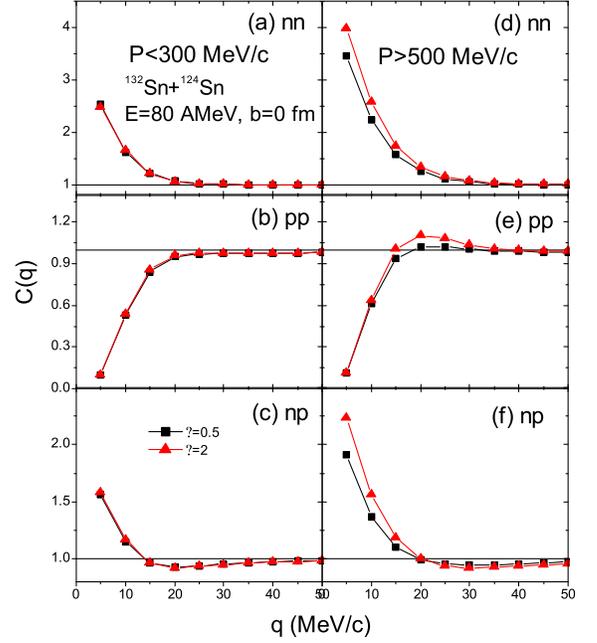} \vspace{0cm}
\caption{{\protect\small (Color online) Two-nucleon correlation functions
gated on total momentum of nucleon pairs using the soft (filled squares) or
stiff (filled triangles) symmetry energy for the central reaction system }$%
^{132}${\protect\small Sn + }$^{124}${\protect\small Sn at $E$}$%
{\protect\small =80}${\protect\small \ MeV/nucleon. Left panels are for $%
P<300$ MeV/c while right panels are for $P>500$ MeV/c.}}
\label{CFSn}
\end{figure}

To see how the above results change for different reaction systems, we show
in Fig. \ref{CFSn} two-nucleon correlation functions for head-on collisions
of $^{132}$Sn + $^{124}$Sn at $E=80$ \textrm{MeV/nucleon} by using the soft
(squares) or stiff (triangles) symmetry energy. This reaction system has a
similar isospin asymmetry, i.e., $\delta =0.22$ as the reaction system $^{52}
$Ca + $^{48}$Ca studied in the above, and can also be studied in future RIA.
It is seen that two-nucleon correlation functions in this case are similar
to those in $^{52}$Ca + $^{48}$Ca reactions shown in Fig. \ref{CFsym}. The
strength of two-nucleon correlation functions at small relative momenta is,
however, about a factor of 2 smaller than that in the reaction system $^{52}$%
Ca + $^{48}$Ca. This reduction is mainly due to the larger size of the
reaction system $^{132}$Sn + $^{124}$Sn than $^{52}$Ca + $^{48}$Ca. Because
of stronger Coulomb repulsion for protons, the symmetry energy effect on the
correlation functions of neutron-proton pairs and proton-proton pairs is
weaker compared with those in the reaction system $^{52}$Ca + $^{48}$Ca,
i.e., about $17\%$ and $8\%$, respectively, instead of $30\%$ and $20\%$.
However, the symmetry energy effect on the neutron-neutron correlation
function, which is not affected by Coulomb potential, is not much influenced
by the size of reaction system, i.e., it is about $20\%$ and $15\%$ for the
light and heavy reaction systems, respectively. As in collisions of $^{52}$%
Ca + $^{48}$Ca, effects due to different isoscalar potentials and \textsl{N}-%
\textsl{N} cross sections on nucleon-nucleon correlation functions are found
to be small.

\section{Summary and outlook}

\label{summary}

In conclusion, using an isospin-dependent transport model we explore
systematically the isospin effects on nucleon-nucleon correlation functions
in heavy-ion collisions induced by neutron-rich nuclei at intermediate
energies. It is found that the nuclear symmetry energy $E_{\mathrm{sym}%
}(\rho )$ influences significantly both the emission times and the relative
emission sequence of neutrons and protons. A stiffer $E_{\mathrm{sym}}(\rho
) $ leads to a faster and almost simultaneous emission of neutrons and
protons. Consequently, two-nucleon correlation functions, especially for
neutron-proton pairs with high total momentum but low relative momentum, are
stronger for the stiff symmetry energy than the soft one. The correlation
function of neutron-proton pairs gated on higher total momentum but with
smaller relative momentum is, on the other hand, insensitive to the
incompressibility of symmetric nuclear matter \textrm{EOS} and the in-medium 
\textsl{N}-\textsl{N} cross sections. This novel property can be used as a
possible tool to extract useful information about the density-dependence of
nuclear symmetry energy. These results do not change much if only nucleons
emitted in certain directions are considered.

We have also studied the dependence of neutron-proton correlation function
on the impact parameter and incident energy of heavy ion collisions as well
as masses of the reaction system. We find that the symmetry energy effect
becomes weaker with increasing impact parameter and incident energy. Also,
the strength of nucleon-nucleon correlation function is reduced in
collisions of heavier reaction systems as a result of larger nucleon
emission source. For proton-proton and neutron-proton correlation functions,
the symmetry energy effect is further suppressed by the stronger Coulomb
potential in heavier reaction systems. Our results thus suggest that the
correlation function of neutron-proton pairs gated on higher total momentum
but with smaller relative momentum from central or semi-central collisions
induced by lighter neutron-rich nuclei is a sensitive probe to the density
dependence of nuclear symmetry energy. We have further studied the
dependence of symmetry energy effect on the time at which IBUU simulations
are terminated. We find that the symmetry energy effect on nucleon-nucleon
correlation functions begins to appear when the maximum nuclear density is
already below the normal nuclear matter density. The effect addressed in
present study via nucleon-nucleon correlation functions is thus largely
limited to the properties of nuclear symmetry energy at subnormal nuclear
densities, similar to that from studying neutron-skins of radioactive nuclei.

In the present work, we have not included momentum dependence in either the
isoscalar mean-field potential or the symmetry potential. The former may
affect the properties of nuclear emission source as it was shown that a
momentum-dependent mean-field potential reduces nuclear stopping or
increases nuclear transparency \cite{greco99}. With momentum dependence
included in the nuclear symmetry potential, differences between the neutron
and proton potentials are affected not only by the density dependence of
nuclear symmetry energy but also by the magnitude of proton and neutron
momenta \cite{das}. It will be of interest to study how the results obtained
in the present study are modified by the momentum dependence of nuclear mean
field. We plan to carry out such a study in the future.

\begin{acknowledgments}
This paper is based on the work supported in part by the U.S. National
Science Foundation under Grant Nos. PHY-0088934 and PHY-0098805 as well as
the Welch Foundation under Grant No. A-1358. Also, the work of LWC is
supported by the National Natural Science Foundation of China under Grant
No. 10105008, and that of VG is supported by the National Institute of
Nuclear Physics (INFN) in Italy.
\end{acknowledgments}

\end{document}